\documentclass[12pt]{article}
\usepackage{amsmath,amssymb,mathtools,amsfonts,epsfig,graphicx,euscript}
 \usepackage{color}
 \usepackage{etoolbox}
 
 \usepackage{xcolor}
 \definecolor{greenblue}{rgb}{0.0, 0.7, 0.1}
  
 \usepackage[colorlinks = true,
 linkcolor = blue,
 urlcolor  = blue,
 citecolor = brown,
 anchorcolor = blue]{hyperref}
 
\begin{document}
\begin{titlepage}
\thispagestyle{empty}

\vspace{4cm}
\begin{center}
\font\titlerm=cmr10 scaled\magstep4
\font\titlei=cmmi10
scaled\magstep4 \font\titleis=cmmi7 scaled\magstep4 {\Large{\textbf{Electroweak phase transition in the presence of hypermagnetic field and the generation of gravitational waves}
\\}}
\setcounter{footnote}{0}
\vspace{1.5cm} 
\noindent{{H. Abedi$^a$ \footnote{e-mail: hamid\_abedi@sbu.ac.ir}
		M. Ahmadvand$^b$ \footnote{e-mail: ahmadvand@ipm.ir} and S. S. Gousheh$^a$ \footnote{e-mail: ss-gousheh@sbu.ac.ir}
		}}\\
\vspace{0.2cm}

{\it $^a$Department of Physics, Shahid Beheshti University, Tehran, Iran\\}
{\it $^b$School of Particles and Accelerators, Institute Research in Fundamental Sciences (IPM), P. O. Box 19395-5531, Tehran, Iran\\}

\vspace*{.4cm}
\end{center}

\vskip 2em
\setcounter{footnote}{0}
\begin{abstract}
We investigate the effects of a large-scale background hypermagnetic field on the electroweak phase transition.  We propose a model in which an effective weak angle  varies during the electroweak phase transition and upon its use we show that, although for the majority of the parameter space the phase transition is a crossover, there are tiny regions in which the phase transition occurs in two steps and can be first-order. We obtain all of the important quantities characterizing the details of the first-order phase transition, including the latent heat, transition temperature and duration. We then calculate the gravitational wave energy spectrum generated during the first-order part of the electroweak phase transition and find that, for strong enough background hypermagnetic fields, these signals can be detected by the Ultimate-DECIGO interferometer.

\end{abstract}
\end{titlepage}

	
	\section{Introduction}
	\label{sec:intro}
	
	Large-scale magnetic fields are ubiquitous in the Universe. The presence of these magnetic fields in galaxies and intergalactic spaces raises the question about their origin \cite{Neronov:1900zz}. One possibility is that these fields have been produced by some astrophysical and dynamo processes \cite{Brandenburg:2004jv}. However, using these mechanisms it is hard to explain the scale of magnetic fields observed in the intergalactic spaces. The other scenario is that these large-scale magnetic fields can be relics of the early Universe. In this case, several mechanisms for magnetogenesis have been proposed to be taking place during the early times such as inflation and cosmological phase transitions \cite{Turner:1987bw}.
	
	
	Considering such large-scale primordial magnetic fields, one can study their effects on significant events taking place after the big bang, e.g.\ the Electroweak Phase Transition (EWPT).
	Assuming they are produced before the EWPT, the large-scale hypermagnetic fields can survive in the plasma with high temperature and conductivity. Due to the chiral Abelian anomaly, these hypermagnetic fields can play an important role in baryogenesis scenarios before the EWPT \cite{Abbaslu:2019yiy, Abbaslu:2020xfn} or during it \cite{Giovannini:1997gp,Giovannini:2015zwr}. Furthermore, a background hypermagnetic field, $B^{\mathrm{bg}}_Y$, can influence the nature of EWPT. Without such a field, the Standard Model (SM) predicts that the EWPT is a crossover. However, the presence of sufficiently strong hypermagnetic fields could change the situation and make it first-order \cite{Giovannini:1997eg, Elmfors:1998wz}. On the other hand, due to the coupling between these fields and magnetic dipole moment of sphalerons, the energy barrier of sphalerons decreases so that sphaleron processes can threaten the EW baryogenesis scenarios in this context \cite{Comelli:1999gt}. In a previous work \cite{Abedi:2018top}, based on gravitational anomaly and chiral gravitational waves (GWs) sourced by helical magnetic fields, we found the possibility to violate $ B-L $ symmetry, where $ B $ and $ L $ are baryon and lepton numbers, respectively. Then, relying on sphaleron processes, we presented a possible mechanism for generation of the matter-antimatter asymmetry of the Universe.
	
	The influence of hypermagnetic fields on the EWPT has initially been addressed in \cite{Giovannini:1997eg} and \cite{Elmfors:1998wz}. Considering only the direct effect of hypermagnetic fields on the Gibbs free energy, they showed analytically that for small enough temperature-dependent Higgs masses \cite{Quiros:1999jp} during the PT and strong enough hypermagnetic fields up to a critical value, the EWPT becomes first-order. The constraints on the Higgs mass and the upper bound of the hypermagnetic field are set to prevent the vorticity problem, i.e. the penetration of Z-component of the hypermagnetic field inside the bubbles, and to prevent the polarization of W fields and formation of W-condensation \cite{Ambjorn:1989sz}, respectively, both of which make the new phase bubbles unstable. These bounds were subsequently modified numerically by lattice simulation \cite{Kajantie:1998rz}. Therefore, the major obstacles for models of first order EWPT in the presence of a background hypermagnetic field has been the vorticity and W-condensation, which we shall address here.
	
	In this work, we start with constructing an appropriate Lagrangian containing the effects of a constant $B^{\mathrm{bg}}_Y$ and calculate an effective potential for the symmetric and broken phases. We then replace the simplifying assumption of a discontinuous change in the weak angle during EWPT with a more realistic assumption of a gradual variation from zero in the symmetric phase to its final value in the broken phase. We show that by considering an effective weak angle which varies during the PT \cite{Kajantie:1996qd}, one can describe the PT analytically.  By analyzing this effective potential we find that as the temperature falls below a temperature $T_0$, the vacuum of the symmetric phase becomes unstable and the usual EW crossover transition to a stable vacuum occurs. However, for some small regions of the model's parameter space, as the temperature decreases further, the presence of a background hypermagnetic field produces a second vacuum whose stability eventually surpasses the first. Consequently, a first-order phase transition can occur between these two well-separated vacua by choosing appropriate subregion of parameter space of the model which satisfies the necessary criteria for avoiding vorticity and W-condensation. Then, by calculating the solution of the bounce equation obtained from the bounce action, we find the transition temperature at which the bubbles nucleate. At this temperature, the vacuum energy of the PT and its relation to the strength of the hypermagnetic field are obtained. We also calculate the duration of the PT. Here, we use the exact expression for the thermal correction part of the one loop effective action, and compare our results to those obtained by the usual high temperature expansion.
	
	A first-order EWPT can have many profound consequences, in particular for matter-antimatter asymmetry generation, as well as for the generation of gravitational waves (GWs). In this paper, we concentrate on the latter. The GWs are useful probes, providing valuable information about the early Universe, partly because they have the least attenuation during their propagation. In general, any first-order PT in the early Universe is regarded as a possible source for the generation of GWs\footnote{ The effect of hypermagnetic fields on GW radiation at the EWPT was first discussed in \cite{Giovannini:1999wv}.} \cite{Cai:2017tmh, Ahmadvand:2017xrw}. In this case, during the evolution of bubbles, three sources can contribute to the production of the GW spectrum: bubble collisions \cite{Kosowsky:1992rz}, magnetohydrodynamic (MHD) turbulence \cite{Kosowsky:2001xp}, and sound waves \cite{Hindmarsh:2013xza}. In this paper, by calculating the characteristic quantities associated with the PTs, we find the GW energy spectrum. We show that the generated GWs can be in the sensitivity range of the future space-based GW experiment, i.e., The Deci-Hertz interferometer Gravitational Wave Observatory (DECIGO), whose the primary objective is to track primordial GWs \cite{Yagi:2011wg,Cooray:2005xr}.
	
	In Section 2, we introduce the one-loop effective action in the presence of a $B^{\mathrm{bg}}_Y$ and obtain the details of the PT including the bounce solution, transition temperature, duration of the PT and latent heat. In Section 3, we calculate the GWs spectra using the parameters obtained in Section 2. We conclude in Section 4. 
	
	
	\section{The electroweak phase transition}
	\subsection{Model}
	The presence of a large-scale background hypermagnetic field, $B^{\mathrm{bg}}_Y$, at the EWPT, regardless of its origin, can have important consequences on the dynamics on the PT. To study these effects, one can decompose the total hypercharge field strength as $f^{{\mathrm{total}}}_{\mu\nu}=f_{\mu\nu} +f^{\mathrm{bg}}_{\mu\nu}  $ and consequently the Lagrangian can be written as 
	\begin{equation}\label{lag}
	\mathcal{L}=\mathcal{L}_{\mathrm{SM}}-\frac{1}{2}(f^{\mathrm{bg}})_{\mu\nu}f^{\mu\nu}-\frac{1}{4}(f^{\mathrm{bg}})_{\mu\nu}(f^{\mathrm{bg}})^{\mu\nu},
	\end{equation}
	where $ \mathcal{L}_{\mathrm{SM}} $ is the Lagrangian of the standard model. As shown in  \cite{Elmfors:1998wz}, one can solve the equations of motion for both the symmetric and broken phases and then obtain the corresponding Gibbs free energies in terms of $B^{\mathrm{bg}}_Y$. 
	
    Conductivity of the primordial plasma is proportional to its temperature \cite{Kajantie:1998rz}, and hence is very high at the EWPT. In such a plasma,
	the only long-range or background field that can survive is the hypermagnetic field, while other fields are either screened, or have temperature induced effective masses, or both. We assume such long-range $B^{\mathrm{bg}}_Y$ exists and, in the case of a first-order PT with bubbles of true vacuum present, its scale is much larger than the typical size of bubbles during the PT. Therefore we take this field to be constant on the scale of the bubbles and further assume $B^{\mathrm{bg}}_Y=bT^2 $ where $b$ is approximately constant during the EWPT.
	However, in order for the Z-component of $B^{\mathrm{bg}}_Y$ not to penetrate the bubbles, the wall width should be larger than the correlation length of the Z-field, i.e., $ l_{\mathrm{wall}}>l_{\mathrm{Z}} $. Also, there is a criterion to avoid the W-condensation, i.e., $eB^{\mathrm{bg}}_Y < (m_{\mathrm{W}} \phi/v)^2 $  \cite{Elmfors:1998wz} where $v$ is the VEV at zero temperature. We shall investigate these constraints in the next section.
	
    The Gibbs free energies corresponding to the effective potentials in the symmetric and broken phases are given by $ V(0, T) $ and $ V(\phi, T)+1/2( B^{\mathrm{bg}}_Y)^2 \sin ^2\theta_{\mathrm{w}} $, respectively. Here, $ \theta_{\mathrm{w}} $ is the Weinberg or weak mixing angle. 
	These free energies are the same as those of \cite{Elmfors:1998wz}, except for a shift of $1/2(B^{\mathrm{bg}}_Y)^2 $ due to the last term of Eq.\ (\ref{lag}). In the absence of an exact theory for the PT based on first principles, one can,
	similar to the Meissner effect, study effects of the energy difference between two phases, i.e., $1/2(B^{\mathrm{bg}}_Y)^2 \sin ^2\theta_{\mathrm{w}} $, by a single form effective potential. 
	
	To obtain this potential, one can consider an effective weak angle which varies during the EWPT from zero in the symmetric phase to its final nonzero value, i.e., $\theta_{\mathrm{w}}$, in the broken phase. Various schemes for the weak angle varying with time have been studied analytically at one-loop level in \cite{Kajantie:1996qd} and by nonperturbative numerical lattice simulations in \cite{DOnofrio:2015gop}. Although the results of analytic and  numerical studies for time dependence of the weak mixing angle agree only marginally, a smooth step-like behavior can be inferred from both studies. Neither of these two studies included a $B_{Y}^{\mathrm{bg}}$. A phenomenological parameterization for $\theta_{\mathrm{w}}$ as a function of temperature for a problem which includes a $B_{Y}^{\mathrm{bg}}$, has been introduced in \cite{Kamada:2016cnb} with a smooth step-like behavior\footnote{The varying mixing angle is expressed as \cite{Kamada:2016cnb}
				\begin{equation}\nonumber
					\cos ^2\theta_{\mathrm{w}}(T)=\cos ^2\theta_{\mathrm{w0}}+\frac{1-\cos ^2\theta_{\mathrm{w0}}}{2} \left[1+\tanh\left( \frac{T-T_{\mathrm{step}}}{\Delta T}\right)\right]
				\end{equation}
				where parameters $T_{\mathrm{step}}$ and $\Delta T$ determine the shape of the step.}.In fact, the gradual variation of the mixing angle during the PT is an essential consequence of the non-Abelian SU(2) gauge fields acquiring mass, which leads to the screening of the isomagnetic fields \cite{Kamada:2016cnb}. In any generic phase transition, the variations of all physical quantities can be naturally regarded as functions of the PT order parameter. In the case of the EWPT the order parameter is the Higgs field vacuum expectation value (VEV) which also varies with time or, equivalently, temperature during the PT. Thus, for our problem which also includes a $B_{Y}^{\mathrm{bg}}$, we propose a phenomenological Ansatz for an effective weak angle as a function of the classical Higgs filed, $\phi$, which in particular, describes the conversion of the $B_{Y}^{\mathrm{bg}}$ into the background magnetic field during the PT gradually and smoothly, as below, \\
		\begin{equation}\label{taeff}
			\Theta_{\mathrm{w}}(\phi) = \theta_{\mathrm{w}} \left[\frac{1}{2}+\frac{1}{2}\tanh\left(\frac{\phi - m}{s}\right)\right].
		\end{equation}
		Here $m$ and $s$ are two free parameters which characterize the midpoint and maximum rate of change of $\Theta_{\mathrm{w}}$ profile, respectively. 
		The parameters $m$ and $s$ are analogous to $T_{\mathrm{step}}$ and $\Delta T$ in the previous model \cite{Kamada:2016cnb}, respectively.
		This profile is a natural form for the weak mixing angle, which depends on the order parameter and varies from zero in the symmetric phase to its value in the broken phase during the phase transition. The free parameters of the effective mixing angle, $m$ and $s$, along with the hypermagnetic field strength, are to be constrained by the requirements necessary for a stable first order PT, i.e., preventing Z-field vorticity and W-field condensation.\\
	
\begin{figure}
		\begin{center}
			\includegraphics[scale=.75]{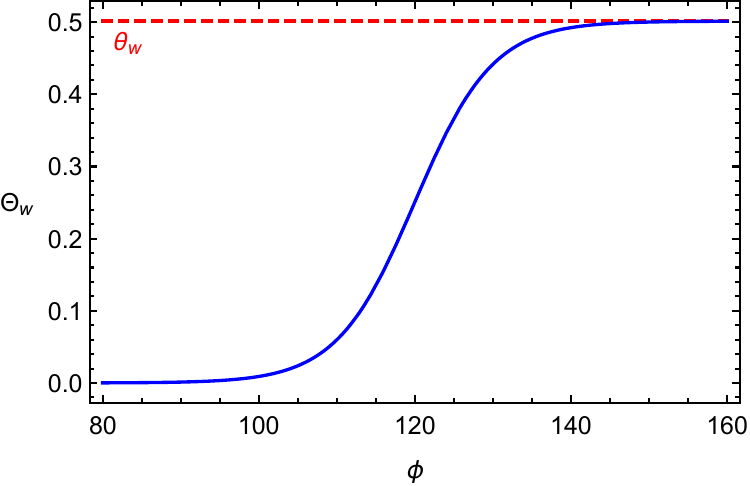}\caption{\label{fan} Our proposed model for the weak angle as a function of $ \phi $ during the EWPT for $m$ = 120 and $s$ = 10.}
		\end{center}
	\end{figure}

	We use the following potential as the finite temperature effective potential  (FTEP) which describes the system during the phase transition and is valid in the symmetric and broken phases
	\begin{equation}\label{eff}
	V_{\mathrm{eff}}(\phi, T,B^{\mathrm{bg}}_Y)=V(\phi, T)+\frac{(B^{\mathrm{bg}}_Y)^2}{2}\sin ^2\Theta_{\mathrm{w}}(\phi).
	\end{equation}
	With the phenomenological Ansatz we have chosen for $\Theta_{\mathrm{w}}(\phi)$, our proposed FTEP coincides with the corresponding Gibbs free energies in the symmetric and broken phases, and can describe the dynamics of the PT continuously. In Fig.\ (\ref{fan}), $\Theta_{\mathrm{w}} $ is shown for $m$ = 120 and $s$ = 10.	
    As we shall show, the form of the PT is the famous crossover for almost the entire parameter space volume. However, we shall demonstrate a two-step PT for some regions of the effective weak angle parameter space.

In Eq.\ (\ref{eff}), $ V(\phi, T) $ consists of the following terms
	\begin{equation}
	V(\phi, T)=V_0(\phi)+V_{1}(\phi)+V_{\mathrm{th}}(\phi,T),
	\end{equation}
	where $ V_0 $, $V_1$, and $V_{\mathrm{th}}$ denote the tree level potential of the Higgs field, the one-loop quantum correction,  and the thermal corrections, respectively. $ V_0 $ is given by
	\begin{equation}
	V_0(\phi)=-\frac{1}{2}\mu ^2\phi ^2+\frac{\lambda}{4}\phi ^4,
	\end{equation}
	where $ \lambda =\mu ^2/\nu ^2 $ is fixed by the Higgs mass, $ \sqrt{2\lambda \nu ^2}=125\,\mathrm{GeV} $, and Higgs VEV at zero temperature, $ \nu =246\,\mathrm{GeV} $. One-loop quantum correction can be written as \cite{Quiros:1999jp}
	
	\begin{eqnarray}
	V_1(\phi)&=&\frac{6}{64\pi ^2}\Big[m_{\mathrm{W}}^4(\phi)\Big(\log \frac{m_{\mathrm{W}}^2(\phi)}{m_{\mathrm{W}}^2(\nu)}-\frac{5}{6}\Big)+2m_{\mathrm{W}}^2(\phi)m_{\mathrm{W}}^2(\nu)\Big]\nonumber\\&+&\frac{3}{64\pi ^2}\Big[m_{\mathrm{Z}}^4(\phi)\Big(\log \frac{m_{\mathrm{Z}}^2(\phi)}{m_{\mathrm{Z}}^2(\nu)}-\frac{5}{6}\Big)+2m_{\mathrm{Z}}^2(\phi)m_{\mathrm{Z}}^2(\nu)\Big]\nonumber\\&-&\frac{12}{64\pi ^2}\Big[m_{\mathrm{t}}^4(\phi)\Big(\log \frac{m_{\mathrm{t}}^2(\phi)}{m_{\mathrm{t}}^2(\nu)}-\frac{3}{2}\Big)+2m_{\mathrm{t}}^2(\phi)m_{\mathrm{t}}^2(\nu)\Big], 
	\end{eqnarray}
	where $ m_{\mathrm{W}}(\phi)=g_2\phi/2 $, $ m_{\mathrm{Z}}(\phi)=\sqrt{(g_2^2+g'^2)}\phi/2 $ and $ m_{\mathrm{t}}(\phi)=y_{\mathrm{t}}\phi /\sqrt{2} $ are masses of the gauge fields and the top quark, respectively. The thermal correction term is as follows \cite{Quiros:1999jp}
	
	\begin{equation}
	V_{\mathrm{th}}(\phi, T)\equiv\sum_{i=W,Z,t}\pm \frac{n_i T^4}{2\pi ^2}J_{\mathrm{B,F}}\Big(\frac{m^2_i(\phi)}{T^2}\Big),
	\end{equation}
	where $ n_i $ is the number of degrees of freedom of the {\it i}\,th  particle and
	\begin{equation}\label{Jexact}
	J_{\mathrm{B,F}}(x)=\int_{0}^{\infty}dy~y^2\log\Big[1\mp\exp\Big(-\sqrt{y^2+x}\Big)\Big].
	\end{equation}
	We solve these integrals exactly, taking the advantage of Cosmo-transition numerical package \cite{Wainwright:2011kj}, without having to use the usual high temperature expansion of these functions, given by
	\begin{equation}\label{thef}
		J_{\mathrm{B}}(x)=\frac{\pi ^2}{12}x-\frac{\pi}{6}x^{\frac{3}{2}}-\frac{x ^2}{32}\log\frac{x}{a_{\mathrm{b}}}+\mathcal{O}(x^3), 
	\end{equation}
	\begin{equation}\label{JF}
		J_{\mathrm{F}}(x)=-\frac{\pi ^2}{24}x-\frac{x ^2}{32}\log\frac{x}{a_{\mathrm{f}}}+\mathcal{O}(x^3), 
	\end{equation}
	where $ \log a_{\mathrm{b}} =5.4076 $ and $ \log a_{\mathrm{f}} =2.6351 $.
	
\begin{figure}
		\begin{center}
			\includegraphics[scale=0.52]{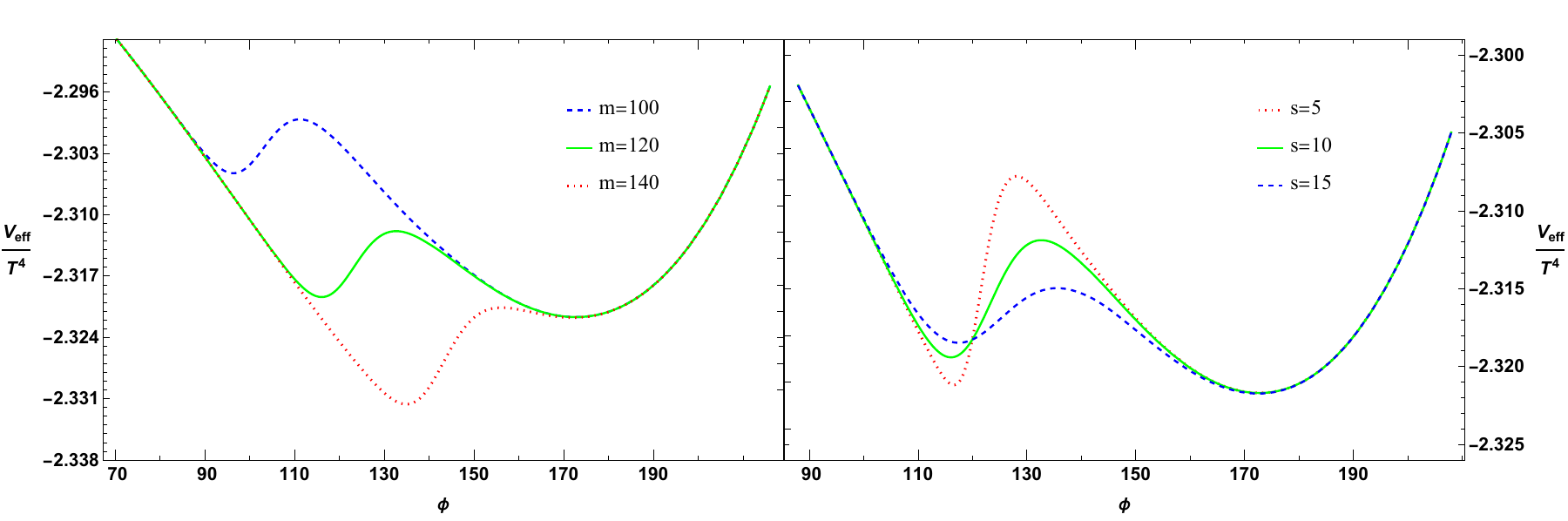}\caption{\label{fbr} The finite temperature effective potential for various values of parameters $m$ and $s$ in the effective weak angle, for $B^{\mathrm{bg}}_Y=0.45\, T^2$ and $T=128.8  GeV$. Left: variation with respect to $m$ at constant $s=10$. Right: variation with respect to $s$ at constant $m=120$.}
		\end{center}
	\end{figure}

\begin{figure}
		\begin{center}
			\includegraphics[scale=0.8]{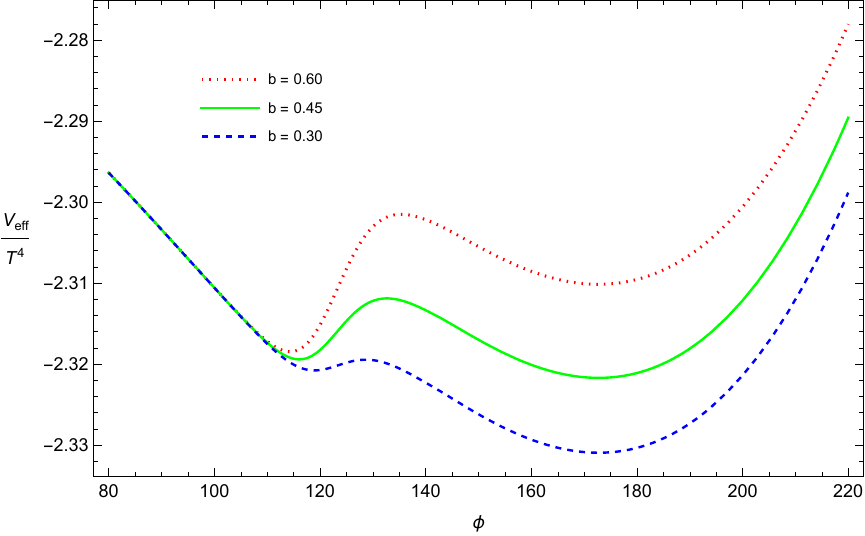}\caption{\label{varT} The finite temperature effective potential for $m=120$, $s=10$ and various values of hypermagnetic field strength, at the nucleation temperature of the green-solid curve, $T=128.8 GeV$.}
				\end{center}
	\end{figure}

	In Fig.\ (\ref{fbr}) FTEP is displayed for different sets of the weak angle parameters. As can be seen from Fig.\ (\ref{fbr}), $m$  affects the transition temperature, the value of the false vacuum and the value of FTEP there, while $s$ mainly affects the height of the barrier. Figure (\ref{varT}) shows the variations of FTEP with respect to the parameter of the hypermagnetic field strength, $b$. As $b$ increases, the phase transition is delayed and becomes stronger. As we shall show, there is an even smaller subset of the parameter space where it is possible to not only avoid the Z-penetration of the bubbles and W-condensation, but also make EW baryogenesis possible.
	
	\subsection{Parameter space of the model}
	In principle, the parameter $s$ can take any arbitrary value. For $s>30$ the PT will be a crossover and for $s<30$ there is a chance for a first order PT. The parameter $m$ must be less than the true vacuum at the end of the PT, around 200. The parameter $b$ has an upper bound around 1 due to the observational restrictions, \cite{Barrow:1997mj,Grasso:1996kk}. However, there is a more restrictive constraint on b from the W-condensation as mentioned before. Free choices of these values in vast majority of the parameter space volume, lead to a crossover. However, as we shall show, for some small subspace of the parameter space the necessary conditions for a first order PT can be fulfilled. 
	
	In order to avoid the vorticity problem, we have to choose the parameters of the model, {\it i.e.}, $m$, $s$ and $b$, in such a way that the resulting bubbles of the true vacuum have the property $l_\mathrm{wall}> 1/m_\mathrm{Z}$, where $l_\mathrm{wall}$ denotes the width of the bubbles wall. Often times in the literature, one uses the estimate $l_\mathrm{wall}\approx 1/m_\mathrm{H}$, in terms of which the vorticity condition becomes $m_\mathrm{H}<m_\mathrm{Z}$. As we shall show, $l_\mathrm{wall}$ is not very sensitive to $m_\mathrm{H}$, and hence the above estimate is not very accurate. We can obtain $l_\mathrm{wall}$ when we compute the bubbles profiles.  The Higgs field mass is obtained directly from the second derivative of $V_{\mathrm{eff}}(\phi, T,B^{\mathrm{bg}}_Y)$ at the location of the true vacuum \cite{Quiros:1999jp}. In our model, in addition to the temperature, it depends on the parameters $m$, $s$ and $b$; but is more sensitive to variations of the parameter $s$: decreasing $s$ makes the barrier of $V_{\mathrm{eff}}(\phi, T,B^{\mathrm{bg}}_Y)$ sharper and increases the Higgs mass. The masses of the other fields in the broken phase are proportional to the Higgs field VEV which in our model is more sensitive to the parameter $m$. These conditions give us an opportunity to pass the vorticity criterion by choosing $s>5$ and $m< 150$. In addition, the existence of a barrier which is the characteristic of a first-order PT imposes an upper bound on $s$ around 30.	Moreover, to overcome the W-condensation problem, the inequality $eB^{\mathrm{bg}}_Y < (m_W \phi/v)^2=g_2^2 \phi ^2/4 $ must be satisfied, which at the transition temperature, $T_*$, where the bubbles of the true vacuum nucleate,  becomes $eB^{\mathrm{bg}}_Y <g_2^2 v(T_*)^2/4 $ or $b <[g_2 v(T_*)/2\sqrt{e}T_*]^2 =: b_\mathrm{max}$.
	
	Also, due to the imprint of primordial magnetic fields on the CMB spectra, the amplitude of these magnetic fields are constrained by Planck data \cite{Ade:2015cva}. With the inverse cascade evolution, the magnitude and correlation length of the magnetic fields at the EW epoch can be related to their values at present \cite{Fujita:2016igl}. The magnitudes for $ B^{\mathrm{bg}}_Y $ at the EW transition with $b\lesssim 1$ are consistent with Planck constraints.
	
	On the other hand, EW baryogenesis requires that the sphaleron rate, which depends on $\phi/T$, should not be suppressed outside the bubbles. This fact sets a lower bound on the parameter $m$.	In order for the sphalerons to be active outside the bubbles before the transition temperature, $v(T)/T$ must be sufficiently less than one. This condition can be satisfied for $ m<120$. On the other hand, the condition $ m>90$ must be satisfied in order to avoid W-condensation. Hence, considering these two constraints, $90<m<120$ is appropriate. Although the domain $m>120$ is not appropriate for EW baryogenesis, we shall include it in our investigation, since it leads to stronger GWs and, after all, the baryogenesis scenarios before EWPT continue to exist as viable alternatives.
	
	Considering all above constraints, we choose the following cubical subregion of the parameter space of the model: $90<m<150$, $5<s<30$, $0.25<b<0.65$. Hereafter, we consider typical representative points in this region as: $b=0.3$, $b=0.45$ and $b=0.6$; $m=100$, $m=120$ and $m=140$; $s=5$, $s=10$ and $s=15$.
	In the next subsection, we calculate all of the major physical quantities characterizing the phase transition.
	
	\subsection{Dynamics of the phase transition}
\begin{figure}
		\begin{center}
			\includegraphics[scale=.9]{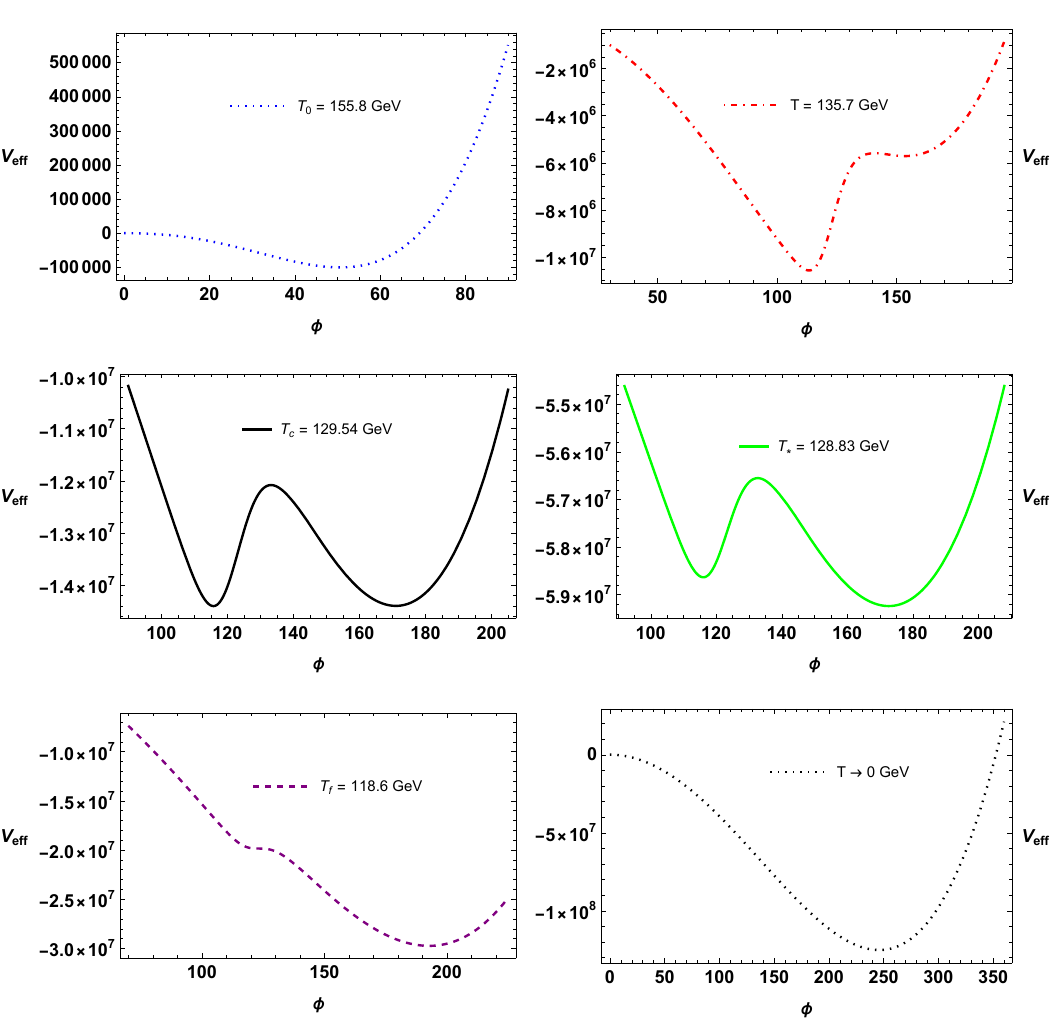}\caption{\label{fef} The finite temperature effective potential is displayed at six different values of temperatures, for $b=0.45$, $m=120$ and $s=10$. The blue-dotted curve at $ T_0=155.8\,\mathrm{GeV} $ shows the onset of crossover, the green dot-dashed curve at $ T=135.7\,\mathrm{GeV} $ depicts the threshold where the second local minimum is forming, the black-solid curve at the critical temperature at $ T_c =129.54\,\mathrm{GeV}  $ shows the presence of two degenerate minima, the red-solid curve corresponds to the nucleation temperature $ T_*=128.83\,\mathrm{GeV}$, the purple-dashed curve at $ T_{\mathrm{f}}=118.6\,\mathrm{GeV} $ shows the threshold where the first minimum disappears and only the second one remains, and the bottom-right curve shows the true vacuum, 246 $\mathrm{GeV}$, when temperature approaches zero.}
		\end{center}
	\end{figure}

	In a typical PT, the existence of a barrier between two minima of the effective potential produces a first-order PT. In \cite{Giovannini:1997eg, Elmfors:1998wz}, using only the difference of the Gibbs free energies of two phases, it is argued on general grounds that the presence of a strong enough $B^{\mathrm{bg}}_Y$ in the effective potential increases the area under the barrier and delays the PT. Here, we use $ V_{\mathrm{eff}} $ as given by Eqs.\ (\ref{taeff}, \ref{eff}) to calculate the dynamics of the EWPT, and show explicitly that this is indeed the case. Before we embark on the calculations, we believe it is essential to first illustrate qualitatively the evolution of FTEP, especially the appearance of a two step phase transition, as temperature decreases through the EWPT. In Fig.\ (\ref{fef}) we plot $V_{\mathrm{eff}}(\phi,T)$ as a function of $\phi$ for a few important temperatures, with the following choice of parameters: $m=120$, $s=10$, and $b=0.45$. In the symmetric phase and well above the EWPT, the system has a stable vacuum at $\phi = 0$. As the temperature decreases to $ T_0 \approx 156\,\mathrm{GeV} $, the second derivative at $\phi = 0$ goes to zero and the vacuum becomes metastable. As the temperature decreases further, a stable minimum starts to form and the VEV continuously increases from zero through the usual EW crossover. The new feature in the presence of a strong $B^{\mathrm{bg}}_Y$ is the formation of a second minimum, when the temperature decreases further, and in this case at $ T\approx 135.7\,\mathrm{GeV} $. At the critical temperature, $T_c =129.54\,\mathrm{GeV}$, the two vacua become degenerate. Shortly after $ T_c $, at nucleation or transition temperature, $T_*=128.83\,\mathrm{GeV}$, the newly formed vacuum becomes the absolute minimum and tunneling to the true vacuum can be fulfilled. Finally, at temperature $ T_{\mathrm{f}}\approx 119\,\mathrm{GeV} $ the barrier between the minima disappears, i.e., only one vacuum remains. This temperature can be considered as the end of the EWPT. Finally, when the temperature approaches zero the VEV at 246 $\mathrm{GeV}$ is obtained. Notice that the EWPT temperature range in our model is roughly $ 120\,\ {\mathrm{ GeV}}\lesssim T \lesssim 160\,\ {\mathrm{ GeV}}$ with center around $130\,\ {\mathrm{ GeV}}$. This has a center shift to lower temperatures of about $20\, \mathrm{ GeV}$ as compared to the lattice  \cite{DOnofrio:2015gop} and one-loop results \cite{Kajantie:1996qd}, due to the presence of a strong $B_{Y}^{\mathrm{bg}}$ \cite{Elmfors:1998wz}.

\begin{figure}
		\begin{center}
			\includegraphics[scale=.4]{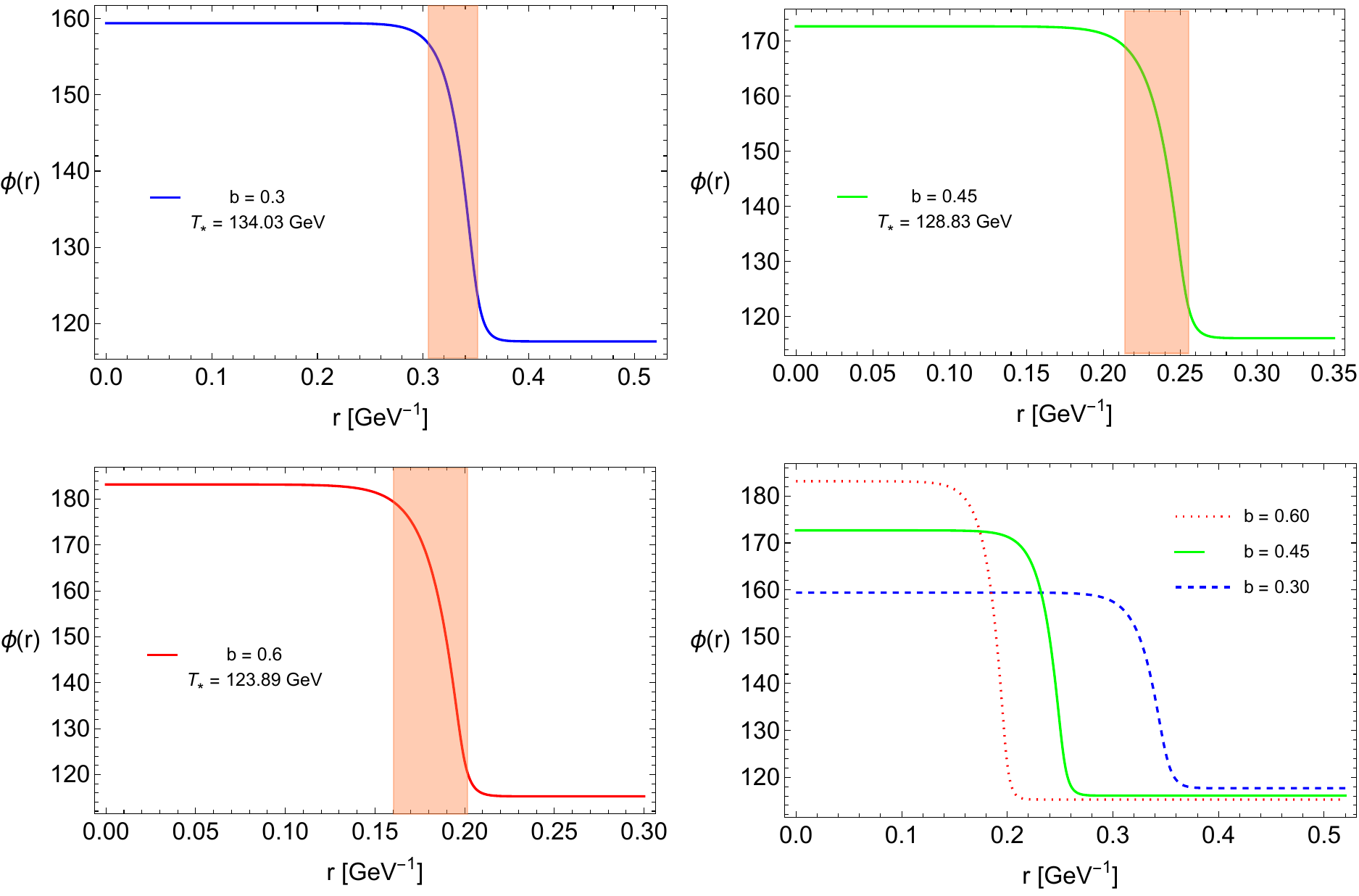}\caption{\label{fbb} The Higgs bubble profiles, also called the bounce solutions, are shown for $m = 120$, $s = 10$ and three different values of $B^{\mathrm{bg}}_Y=bT^{2}_{*}$. The bubble wall width, $l_{\mathrm{wall}}$, is about $ 0.04\,\mathrm{GeV}^{-1}$ in all profiles.}
		\end{center}
	\end{figure}

Having illustrated the general features of the two-step PT, we can proceed to calculate the details of bubble formation between the two vacua for $T<T_c$ which, for the example illustrated in Fig.\ (\ref{fef}), are at $\phi_{\mathrm{false}}\approx 116$ Gev and $\phi_{\mathrm{true}}\approx 172$ GeV. The probability of true vacuum bubble nucleation per unit Hubble space-time volume at finite temperature is given by \cite{Linde:1980tt}
\begin{equation}\label{pro}
P\simeq \frac{ M_{\mathrm{pl}}^4}{T^4}\exp\Big(-\frac{S_3(T)}{T}\Big), 
\end{equation}
where $ M_{\mathrm{pl}} $ is the Planck mass and $ S_3(T) $ is the three-dimensional Euclidean bounce action\footnote[3]{It should also be noted that the hypermagnetic field may deform the bubbles and downgrade the spherical symmetry to an axial symmetry. Therefore, the calculations presented are an approximation. An attempt to improve the calculations should include the back-reaction of the plasma which might produce a restoring force.}
\begin{equation}\label{bac}
S_3(T)=\int_{0}^{\infty}4\pi r^2~dr\Big[\frac{1}{2}\Big(\frac{d\phi}{dr}\Big)^2+V_{\mathrm{eff}}(\phi, T,B^{\mathrm{bg}}_Y)\Big]. 
\end{equation}
As shown below, the presence of a strong hypermagnetic field can significantly affect the bounce action, and in particular the EWPT is delayed. The transition and bubble nucleation occur when the bubble formation probability is of the order of one. From this condition, we can find the transition temperature, $T_* $, \cite{Linde:1980tt}
\begin{equation}\label{bou}
\frac{S_3(T_*)}{T_*}=4\ln\Big(\frac{T_*}{H_*}\Big), 
\end{equation}
where $ H\simeq T^2/ M_{\mathrm{pl}}  $ is the Hubble expansion parameter, and $S_3$ denotes the bounce action which has been minimized as a functional of $\phi$. The latter is sometimes referred to as the on-shell bounce action.  Extremizing the bounce action leads to the bounce equation which is shown below along with the appropriate boundary conditions,
\begin{equation}
\frac{d^2\phi}{dr^2}+\frac{2}{r}\frac{d\phi}{dr}=\frac{\partial V_{\mathrm{eff}} }{\partial\phi},~~~~~~~\left.\frac{d\phi}{dr}\right|_{r=0}=0,~~~~\phi(\infty)=\phi_{\mathrm{false}}.
\end{equation}
From these equations, we can find the radial profile of the Higgs, i.e., the bounce solution, which, as we shall see, connects the two phases through a smooth step-like function. To obtain the solution, we use the \textquotedblleft any bubble\textquotedblright \, code   \cite{Masoumi:2016wot} which takes advantage of a multiple-shooting method. The result is shown in Fig.\ (\ref{fbb}). We can use the following observation as a consistency check on our solutions: the expectation value of the Higgs field inside the bubble for the middle case is completely consistent with $\phi_{\mathrm{true}}=172.7$, which is the absolute minimum of the green solid curve, corresponding to $T=T_*$, for the example depicted in Fig.\ (\ref{fef}). Moreover, as shown in Fig.\ (\ref{fbb}), the wall width of the true vacuum bubbles, $l_{\mathrm{wall}} \gtrsim 0.04\,\mathrm{GeV}^{-1}$, is larger than the correlation length of Z-field, $ l_{\mathrm{Z}}\sim 1/m_{\mathrm{Z}}=2/(\sqrt{g_2^2+g'^2}v(T_*))\sim 0.01-0.02\,\mathrm{GeV}^{-1} $, which can be considered as a recheck for results that satisfy the vorticity criterion. However, for $20<\mathrm{s}<30$ even the inequality $m_\mathrm{H}<m_\mathrm{Z}$ can be fulfilled. Also, to prevent the W-condensation at the transition temperature, the inequality $b<b_\mathrm{max}$ must be satisfied, which sets an upper bound for the values of $ B^{\mathrm{bg}}_Y $ for any choice of $m$ and $s$ in the effective weak angle. Below, we shall calculate and tabulate the values of $l_{\mathrm{wall}},m_{\mathrm{Z}},m_{\mathrm{H}}$ and $b_\mathrm{max} $ for several representative points, showing explicitly that the conditions for the absence of vorticity and W-condensation are satisfied. However, there are important thermodynamic quantities that we have to calculate first, which we address below.
	
\begin{figure}
	\begin{center}
		\includegraphics[scale=.67]{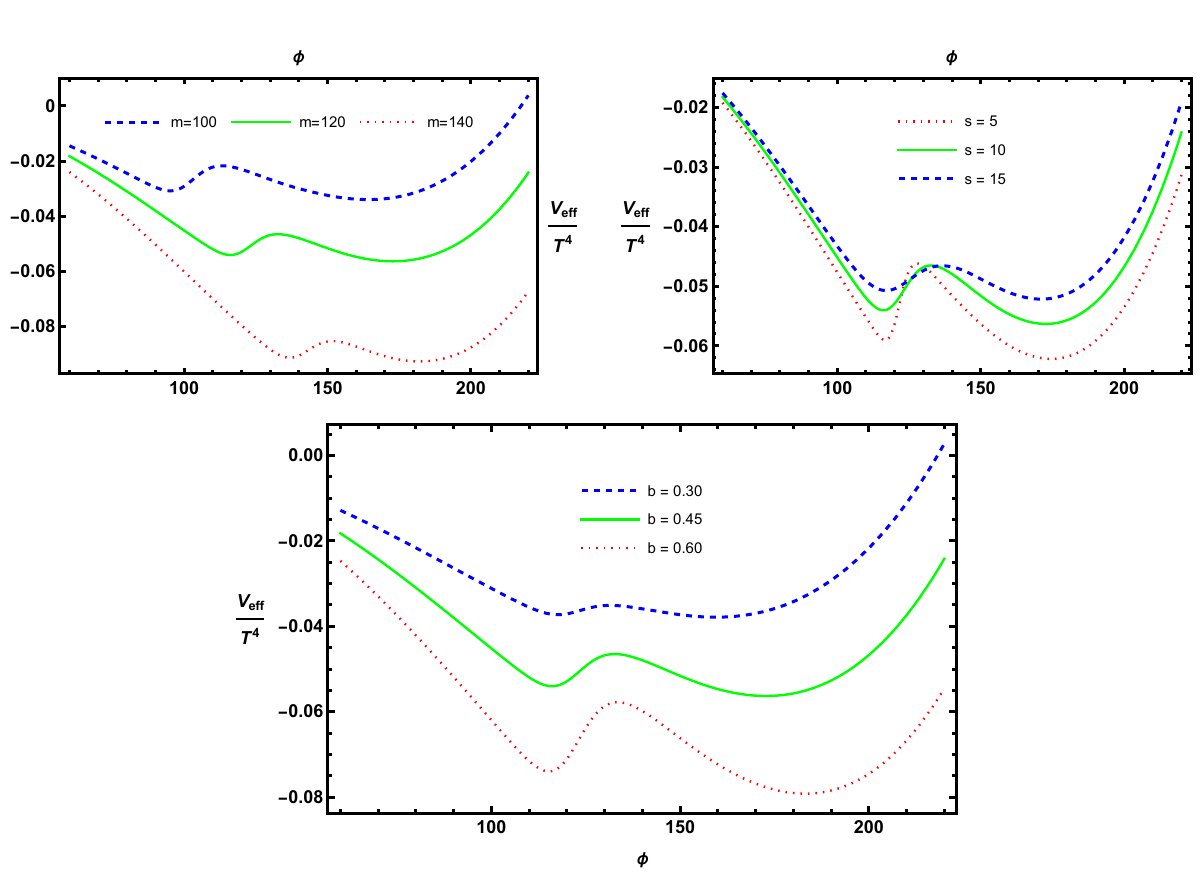}\caption{\label{feb} The finite temperature effective potential is shown for the same set of parameters as in Figs. (2) and (3), but all at their corresponding transition temperatures, as displayed in Table \ref{t1}.}
	\end{center}
\end{figure}

Putting the bounce solution into Eq.\ (\ref{bac}) and computing the integral, we obtain $ S_3(T)/T $ as a function of temperature. Then, using Eq.\ (\ref{bou}), we calculate the transition temperature, $ T_* $, for some typical points in the parameter space and show the results in Fig.\ ({\ref{feb}}). As seen from Fig.\ ({\ref{feb}}), by increasing $B^{\mathrm{bg}}_Y$, the barrier between the vacua is increased and PT occurs at a lower temperature while decreasing the parameter $s$ improves the height of the barrier and variations of parameter $m$ mostly affect the location of the false vacuum.

Other important PT characteristics such as vacuum energy and duration of the PT are computed at $ T_* $.  First, the latent heat at $T_c$ is given by 
	\begin{equation}
	L=-T_c\frac{d\Delta V_{\mathrm{eff}}(T)}{dT}\Bigg|_{T_c}, 
	\end{equation}
	from which one can define the vacuum energy density as
	\begin{equation}\label{eps}
	\epsilon _*=\left(\Delta V_{\mathrm{eff}}(T)-T\frac{d\Delta V_{\mathrm{eff}}(T)}{dT}\right)\Bigg|_{T=T_*},
	\end{equation}
	where $ \Delta V_{\mathrm{eff}}(T_*)=V_{\mathrm{eff}}[\phi_{\mathrm{false}}(T_*), T_*]-V_{\mathrm{eff}}[\phi_{\mathrm{true}} (T_*), T_*] $. Moreover, the duration of the PT, $ \tau ^{-1} $, can be obtained by the following procedure. We assume bubbles nucleate with a rate per space-time volume which is given by\footnote[5]{Here we adhere to the usual definition, with the understanding that $t$ terminates at $t_*$ corresponding to $T_*$.} $ P=P_0\exp(\tau t) $. As a result, we have $ \tau =\dot{P}/P $. Then, using $ dT/dt\simeq -HT $ and Eq.\ (\ref{pro}), we can find $ \tau $ from $ S_3(T)/T $ at $ T_* $ obtained in the previous section
	\begin{equation}
	\frac{\tau}{H_*}=T_*\frac{d}{dT}\Big(\frac{S_3(T)}{T}\Big)\Big |_{T_*}. 
	\end{equation}
	  $\tau/{H_*}$ is a dimensionless quantity which is a measure of the rate of the first-order part of the transition and has significant effects on the energy and frequency distributions of the generated GW \cite{Kamionkowski:1993fg}. In Table \ref{t1}, we show values of various quantities characterizing the EWPT, including $ \epsilon _* $ and $ T_* $, for some representative points in the parameter space. It is interesting to note that the PT duration, $\tau^{-1}\sim 10^{-3}(t_{\mathrm{f}}-t_*)$, where $t_{\mathrm{f}}$ and $t_*$ are the times associated with $T_{\mathrm{f}}$ and $T_*$, respectively. In the next section, we will use these quantities to compute the spectrum of the GWs produced during the EWPT.
	\begin{table}
		\begin{center}
			\begin{tabular}{|c| c| c| c| c| c| c| c| c| c| c| c| c|} 
				\hline
				$b$ & $ m $ & $ s $ & $ T_* $& $ V_f $ & $ V_t $ & $l^{-1}_{\mathrm{wall}}$ & $m_{\mathrm{H}}$ & $m_{\mathrm{Z}}$ & $ b_\mathrm{max} $& $ L $ & $\epsilon _*$ & $\tau /H_*$\\
				\hline\hline
				0.45&120&\textcolor{purple}{\textbf{5}}&128.0&116.8&174.6&25.5&78.5&64.1&0.62&1.13&1.14&20229 \\ 
				\hline
				0.45&120&\textcolor{purple}{\textbf{10}}&128.8&116.1&172.7&23.6&77.3&63.5&0.59&1.09&1.10&22366 \\
				\hline
				0.45&120&\textcolor{purple}{\textbf{15}}&129.4&117.0&171.1&23.3&74.8&62.9&0.58&1.06&1.07&26009 \\
				\hline
				0.45&\textcolor{orange}{\textbf{100}}&10&132.3&95.1&164.2&23.8&72.2&60.3&0.51&1.30&1.32&21686 \\
				\hline
				0.45&\textcolor{orange}{\textbf{120}}&10&128.8&116.1&172.7&23.6&77.3&63.5&0.59&1.09&1.10&22366 \\
				\hline
				0.45&\textcolor{orange}{\textbf{140}}&10&124.4&136.9&182.1&25.5&82.3&66.9&0.73&0.90&0.90&28422 \\
				\hline
				\textcolor{blue}{\textbf{0.30}}&120&10&134.0&117.7&159.4&20.8&68.3&58.6&0.47&0.80&0.80&42918 \\
				\hline
				\textcolor{blue}{\textbf{0.45}}&120&10&128.8&116.1&172.7&23.6&77.3&63.5&0.59&1.09&1.10&22366 \\
				\hline
				\textcolor{blue}{\textbf{0.60}} & 120 & 10 & 123.9 & 115.2 & 183.1 &25.2 &83.5 &67.3 &0.73 & 1.31&1.40&19732 \\
				\hline
			\end{tabular}
			\caption{The values of the transition temperature $T_*$, the false vacuum $V_f$, the true vacuum $V_t$, the inverse of bubble wall width $l^{-1}_{\mathrm{wall}}$, the Higgs mass $m_{\mathrm{H}}$, the Z field mass $m_{\mathrm{Z}}$, the maximum allowed hypermagnetic field strength parameter  $b_\mathrm{max}$, the latent heat $ L $, the vacuum energy density $\epsilon _*$, and the scaled rate of the first order part of the transition $\tau /H_*$ are shown for all representative points specified by various values of parameters $b$, $ m $ and $ s $. The values of $ T_* $, $ V_f $, $ V_t $, $l^{-1}_{\mathrm{wall}}$, $m_{\mathrm{Z}}$ and $m_{\mathrm{H}}$ are in $ \mathrm{GeV} $, and the latent heat, $L$, and $\epsilon_*$ are in $\mathrm{GeV}^4 \times 10^8$.}\label{t1}
		\end{center}
	\end{table}

 Finally and in passing, we like to mention that if we use the more commonly-used high temperature expansion of thermal functions, Eqs.\ (\ref{thef}) and (\ref{JF}), instead of using the exact expression Eq.\ (\ref{Jexact}), we would get different results as shown in Table \ref{t11}. Comparison of the results shows that we obtain a stronger first order phase transition when using the exact expression. In particular, the jump in the order parameter defined by $ \Delta \phi(T_*)=\phi_{\mathrm{true}}(T_*)-\phi_{\mathrm{false}}(T_*)$ increases by about $2\% $ and the latent heat released increases by about $15\% $. Regarding the two step PT, the bubbles of the new phase nucleates within a background that has already acquired some value for the Higgs field VEV. Therefore, the strength of the second step, i.e. the first-order part, can be defined as: $ \Delta \phi(T_c)/T_c$, where $ \Delta \phi(T_c)=\phi_{\mathrm{true}}(T_c)-\phi_{\mathrm{false}}(T_c)$. According to Table \ref{t1}, the values of this quantity lie between 0.31 to 0.56. In the following, we use the results obtained by the exact expression, as shown in Table \ref{t1}.
	\begin{table}
		\begin{center}
			\begin{tabular}{|c| c| c| c| c| c| c|} 
				\hline
				$ b $ & $ m $ & $s$&$ T_* $  &$ V_f $ & $ V_t $&  $ L $ \\
				\hline\hline
				0.3 & 120 & 10 &139.0 & 117.9 & 159.4 &0.69 \\ 
				\hline
				0.45 & 120 &10 &  133.3& 115.9 & 172.1 &0.94 \\
				\hline
				0.6 & 120 &10 & 128.6 &115.3  & 183.0& 1.2 \\
				\hline
			\end{tabular}
			\caption{Same as the last three rows of the Table \ref{t1}, except the thermal correction terms of the one-loop effective action are computed using the high temperature expansions instead of the exact integral expression.}\label{t11} 
		\end{center}
	\end{table}

\section{Gravitational wave generation}
	In this section, we calculate the spectrum of the GWs generated during a first-order EWPT. Considering the ever increasing detection capabilities of GW detectors, GWs can be used as an effective probe of the early Universe. In particular, we investigate whether the GWs produced in our model fall within the detection range of Ultimate-DECIGO.
	
	During cosmological first-order PTs and evolution of the bubbles of true vacuum, three processes can give rise to GW radiation. Indeed, when bubbles nucleate and grow, because of their collisions, part of the latent heat released during the transition is converted to GWs. Moreover, a fraction of the energy is transferred to the plasma and causes the plasma motion which in turn provides two other GW sources: MHD turbulence and sound waves.
	
	The contribution of the first mentioned source to the GW energy density spectrum is calculated by numerical simulations using the envelope approximation and expressed in terms of the PT parameters \cite{Kamionkowski:1993fg}:
	
	\begin{equation}\label{spe}
	h^2\Omega _{\mathrm{col}}(f)=1.67\times 10^{-5}\Big(\frac{0.11 v_{\mathrm{b}}^3}{0.42+v_{\mathrm{b}}^2}\Big) \Big(\frac{H_*}{\tau}\Big)^{2}\Big(\frac{\kappa \alpha}{1+\alpha}\Big)^2 \Big(\frac{100}{g_*}\Big)^{\frac{1}{3}}S_{\mathrm{en}}(f),
	\end{equation}
	where $h$ is the present Hubble parameter $H_0$ in units of 100 km $\mathrm{sec}^{-1}\mathrm{Mpc}^{-1} $, $ v_{\mathrm{b}} $ is the bubble wall velocity, the factor $ \kappa $ stands for the fraction of the vacuum energy which is converted into the kinetic energy of the bubbles, $ g_*\simeq 106 $ is the number of effective relativistic degrees of freedom at the EWPT, and $ \alpha $ denotes the ratio of the vacuum energy density to the thermal energy density,
	\begin{equation}\label{al}
	\alpha =\frac{\epsilon _*}{\frac{\pi^2}{30}g_*T_*^4}.
	\end{equation}
	The Expression for $ \epsilon _* $ is given by Eq.\ (\ref{eps}). The spectral shape of the GWs is given by the following analytic fit \cite{Huber:2008hg}
	\begin{equation}
	S_{\mathrm{en}}(f)=\frac{3.8(\frac{f}{f_{\mathrm{en}}})^{2.8}}{1+2.8(\frac{f}{f_{\mathrm{en}}})^{3.8}},
	\end{equation}
	where the present-day red-shifted peak frequency is given by the following relation,
	\begin{equation}
	f_{\mathrm{en}}=16.5\times 10^{-6}  \Big(\frac{0.62}{1.8-0.1 v_{\mathrm{b}}+v_{\mathrm{b}}^2}\Big)\Big(\frac{\tau}{H_*}\Big)\Big(\frac{T_*}{100~\mathrm{GeV}}\Big)\Big(\frac{g_*}{100}\Big)^{\frac{1}{6}}.
	\end{equation}
	
	Furthermore, we should take into account the other two sources contributing to the GW energy density. Gravitational wave contributions form sound waves, which is numerically calculated in \cite{Hindmarsh:2015qta}, and MHD turbulence as a Kolmogorov-type turbulence modeled by \cite{Caprini:2009yp} are given by
	\begin{equation}\label{sps}
	h^2\Omega _{\mathrm{sw}}(f)=2.65\times 10^{-6}\Big(\frac{H_*}{\tau}\Big)\Big(\frac{\kappa _{\mathrm{sw}} \alpha}{1+\alpha}\Big)^2\Big(\frac{100}{g_*}\Big)^{\frac{1}{3}} v_{\mathrm{b}}~ S_{\mathrm{sw}}(f),
	\end{equation}
	and
	\begin{equation}\label{spt}
	h^2\Omega _{\mathrm{tu}}(f)=3.35\times 10^{-4}\Big(\frac{H_*}{\tau}\Big)\Big(\frac{\kappa _{\mathrm{tu}} \alpha}{1+\alpha}\Big)^{\frac{3}{2}}\Big(\frac{100}{g_*}\Big)^{\frac{1}{3}} v_{\mathrm{b}}~ S_{\mathrm{tu}}(f),
	\end{equation}
	where their spectral shapes are as follows \cite{Caprini:2015zlo},
	\begin{eqnarray}
	S_{\mathrm{sw}}(f)&=&\Big(\frac{f}{f_{\mathrm{sw}}}\Big)^3\Big(\frac{7}{4+3(\frac{f}{f_{\mathrm{sw}}})^{2}}\Big)^{\frac{7}{2}}, \\
	S_{\mathrm{tu}}(f)&=&\frac{(\frac{f}{f_{\mathrm{tu}}})^3}{(1+\frac{f}{f_{\mathrm{tu}}})^{\frac{11}{3}} (1+\frac{8\pi f}{h_*})},
	\end{eqnarray}
	with
	\begin{equation}
	h_*=16.5\times 10^{-6} [\mathrm{Hz}]\Big(\frac{T_*}{100~\mathrm{GeV}}\Big)\Big(\frac{g_*}{100}\Big)^{\frac{1}{6}},
	\end{equation}
	as the red-shifted Hubble parameter. The red-shifted peak frequencies in the spectral shapes of these GW spectra are given by
	\begin{eqnarray}\label{fst}
	f_{\mathrm{sw}}=1.9\times 10^{-5}  \Big(\frac{1}{v_{\mathrm{b}}}\Big)\Big(\frac{\tau}{H_*}\Big)\Big(\frac{T_*}{100~\mathrm{GeV}}\Big)\Big(\frac{g_*}{100}\Big)^{\frac{1}{6}}, \nonumber\\
	f_{\mathrm{tu}}=2.7\times 10^{-5} \Big(\frac{1}{v_{\mathrm{b}}}\Big)\Big(\frac{\tau}{H_*}\Big)\Big(\frac{T_*}{100~\mathrm{GeV}}\Big)\Big(\frac{g_*}{100}\Big)^{\frac{1}{6}}.
	\end{eqnarray}
	
	As can be easily seen from Eqs.\ (\ref{spe}-\ref{fst}), the bubble growth velocity, $ v_{\mathrm{b}}$, has an important role in the energy density spectrum of GWs originating from each of the three sources. An important parameter which affects $ v_{\mathrm{b}}$ is $ \alpha $, as defined in Eq.\ (\ref{al}). In particular, a critical value of  $ \alpha $ is given by \cite{Caprini:2015zlo},
	\begin{equation}\label{inf}
	\alpha_{\infty}=\frac{30}{24 \pi ^2}\frac{\sum _i c_i \Delta m_i ^2(\phi _*)}{g_* T_*^2},
	\end{equation}
	where $ c_i=n_i\,(c_i=n_i/2) $ and $ n_i $ is the number of degrees of freedom for boson (fermion) species, and $ \Delta m_i^2 $ is the squared mass difference of particles between two phases. Thus, the main contribution to $ \alpha_{\infty} $ comes from the particles becoming heavy during the PT, i.e., $ W $, $ Z $ gauge bosons and $ t $ quark.
	It has been argued that the frictional forces due to the surrounding plasma are independent of the Lorentz factor $\gamma=(1- v_{\mathrm{b}}^2)^{-1/2} $ of the bubble wall \cite{Bodeker:2009qy}. Thus, for $ \alpha >\alpha_{\infty} $ the bubbles keep accelerating and run away, resulting in $ v_{\mathrm{b}}=1 $. In this case, all three sources contribute to the GW spectrum, i.e., $ h^2\Omega (f)\simeq h^2\Omega _{\mathrm{col}}+h^2\Omega _{\mathrm{sw}}+h^2\Omega _{\mathrm{tu}} $. However, a more recent study \cite{Bodeker:2017cim} shows that by considering a next-to-leading order calculation, massive gauge bosons develop an additional frictional term proportional to $\gamma$, and hence bubble walls reach a terminal velocity, averting runaway bubbles. This makes the contribution of bubble collisions negligible for generic values of $ \alpha >\alpha_{\infty} $. However, for very strong PTs with $ \alpha\gg 1 $, bubble collisions can contribute significantly to the GW spectrum \cite{Ellis:2019oqb}.
	
	In our case, according to $\alpha$ values listed in Table \ref{t2} for three different values of $B^{\mathrm{bg}}_Y$, the dominant contributions to GWs come from sound waves and MHD turbulence, i.e., $ h^2\Omega (f)\simeq h^2\Omega _{\mathrm{sw}}+h^2\Omega _{\mathrm{tu}} $. For the calculation of bubble wall velocity, we use the Jouguet detonation regime in which $ v_{\mathrm{b}} $ is given by \cite{Espinosa:2010hh} 
	\begin{equation}
	v_{\mathrm{b}}=\frac{\sqrt{\alpha^2+2\alpha/3}+\sqrt{1/3}}{1+\alpha}.
	\end{equation}
	In our case, the efficiency factor for the conversion of the latent heat to the plasma motion can be expressed as \cite{Caprini:2015zlo,Espinosa:2010hh} 
	\begin{eqnarray}
	\kappa _{v}=\frac{\alpha}{0.73+0.083\sqrt{\alpha}+\alpha}.
	\end{eqnarray}
	In addition, the fraction of plasma motion which is turbulence, $\varepsilon=\kappa _{\mathrm{tu}}/\kappa _{\mathrm{v}} $, can be of the order of $ \varepsilon =0.1 $ \cite{Hindmarsh:2015qta}. Therefore, the dominant source coming from the plasma motion is attributed to the sound waves, $ \kappa _{\mathrm{sw}}=(1-\varepsilon)\kappa _{\mathrm{v}} $. \\
	\begin{table}
		\begin{center}
			\begin{tabular}{|c| c| c| c| c| c|} 
				\hline
				$b$ & $m$ & $s$ & $ \alpha $   & $ \alpha _{\infty} $ & $ v_{\mathrm{b}}$  \\
				\hline\hline
				0.30 & 120 &10 & 0.007  & 0.003& 0.64 \\ 
				\hline
				0.45 & 120 & 10 & 0.012  & 0.005& 0.66 \\
				\hline
				0.60 & 120 & 10 & 0.016  & 0.006& 0.67 \\
				\hline
			\end{tabular}
			\caption{The values of $\alpha$, $\alpha_{\infty}$ and $ v_{\mathrm{b}}$, which are necessary to specify GW signals, are displayed for our three representative points, which are detailed in the last three rows of the Table \ref{t1}.}\label{t2}
		\end{center}
	\end{table}
	
\begin{figure}
		\begin{center}
			\includegraphics[scale=.58]{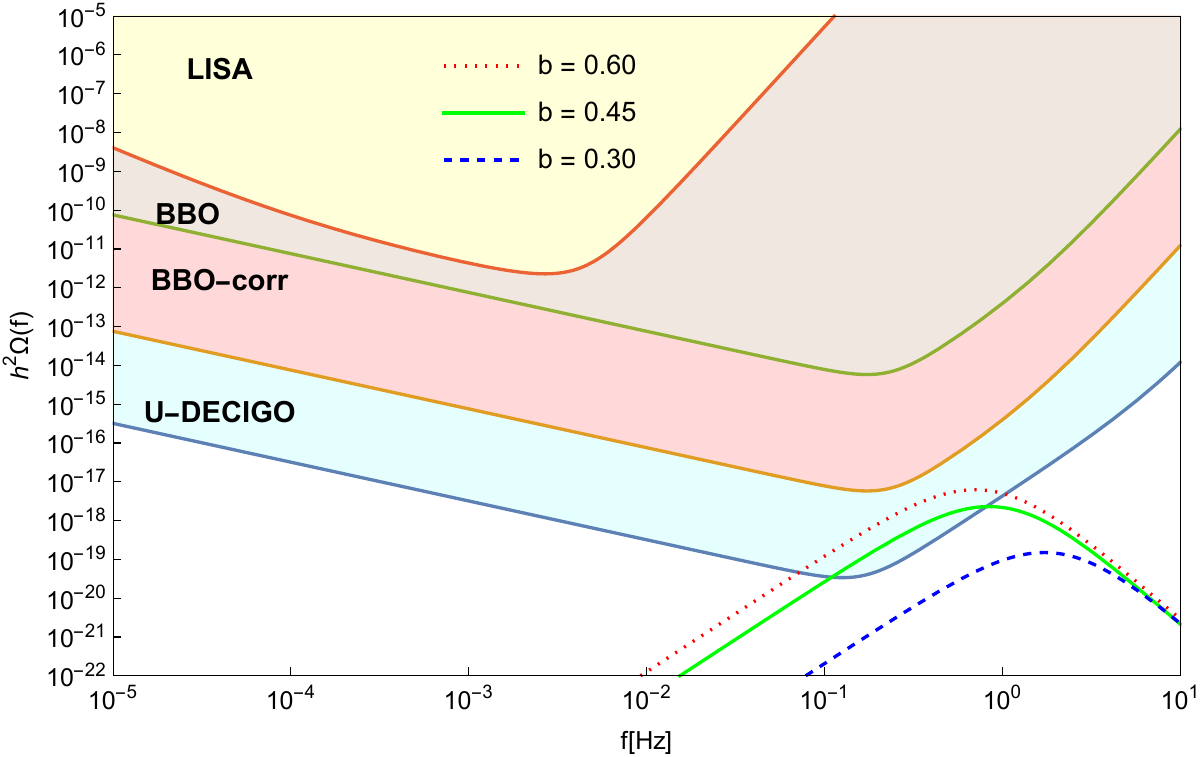}\caption{\label{f2} We display the energy density of GWs generated from the first-order EWPT within our model, for the three representative points shown in the last three rows of Tables \ref{t1} and Table \ref{t2}. Note that as the value of  $B^{\mathrm{bg}}_Y$ increases, the energy density of GWs grows, while the peak frequency decreases. As can be seen, for large enough $B^{\mathrm{bg}}_Y$, these GWs can be in the sensitivity range of the ultimate sensitivity of DECIGO, the  Ultimate-DECIGO interferometer \cite{Yagi:2011wg}, while the signals do not fall within the reach of LISA \cite{Klein:2015hvg}, BBO \cite{Crowder:2005nr} and BBO-correlated \cite{Yagi:2011wg,Crowder:2005nr} detectors.}
		\end{center}
	\end{figure}

	Now, having obtained the key parameters and having determined the contribution of each GW source, Eqs.\ (\ref{sps}) and (\ref{spt}), we compute the GW spectrum generated from the EWPT. In Fig.\ (\ref{f2}), we show the GW spectrum for the three representative points shown in Table \ref{t2}. As listed in Table \ref{t2}, the cases with larger values of $B^{\mathrm{bg}}_Y$ give rise to larger values of $ \alpha $ and so stronger first-order EWPT, leading to GWs with higher energy density but lower peak frequency (see also Fig.\ (\ref{feb})).
	
	As can be seen in Fig.\ (\ref{f2}), these GWs can be detected by the future space-based Ultimate-DECIGO interferometer \cite{Yagi:2011wg}, which will be able to detect GWs around $ 0.1-10\,\mathrm{Hz} $ at which noises raised from irresolvable gravitational wave signals are negligible \cite{Sato:2017dkf}. Therefore, this detector has higher sensitivity and can cover the gap frequency band between LISA and ground-based detectors. Moreover, the interferometer which consist of several detectors can enhance its sensitivity by a few orders of magnitude by making correlation analysis between independent detectors. In fact, Ultimate-DECIGO interferometer can reach a sensitivity level of the order of $ 10^{-20}$ around $ 0.1\,\mathrm{Hz} $ \cite{Cooray:2005xr}. Finally, we expect that these GW spectra predicted by our model with peak frequency between $ 0.1-1\,\mathrm{Hz} $ can be captured by the most sensitive frequency range of the detector, provided $B^{\mathrm{bg}}_Y$ is large enough.
		
	
	\section{Conclusion}
	The finite temperature effective potential in the presence of a background hypermagnetic field is a function of the electroweak mixing angle \cite{Giovannini:1997eg, Elmfors:1998wz}. The mixing angle has been considered as varying continuously during the PT \cite{Kajantie:1996qd, Kamada:2016cnb}. In this work, we have considered it as varying, like the variations of other physical quantities in a PT, as a function of the order parameter. It is shown that the vast majority of the parameter space of the model leads to a crossover. 
	However, we have shown that, for a small subregion of the parameter space, {\it i.e. $s<30$,} the presence of a large-scale background hypermagnetic field, $B^{\mathrm{bg}}_Y$, can have profound effects on the EWPT. In particular, we have shown that as the temperature drops, $B^{\mathrm{bg}}_Y$ can produce a second minimum at a yet larger value of $ \phi $ which eventually becomes the true vacuum. This makes a two-step transition: a crossover to a temporary vacuum followed by a first-order to the final vacuum. The barrier between the two minima also depends on the strength of $B^{\mathrm{bg}}_Y$. We have also shown that there exists an even smaller subregion of the parameter space where we can have a stable first order phase transition in which the conditions for the absence of vorticity and W-condensation are also satisfied. The latter condition requires $b<b_\mathrm{max}(b,m,s)$. Within the range of parameters studied here, $b_\mathrm{max}(b,m,s)$ increases approximately as the square root of $b$, linearly with $m$, and decreases slightly by increasing $s$. For our central case $m=120$ and $s=10$, $b$ can be at most $0.82$, {\it, i.e.} $b_\mathrm{max}(0.82,120,10)\sim 0.82$. Moreover, to avoid the vorticity problem the bubbles of the true vacuum should have the property $l_\mathrm{wall}> 1/m_\mathrm{Z}$, and this can be accomplished in our model by choosing $s>5$ and $m< 150$.  In this regard, we have shown that the usual estimate $l_\mathrm{wall}\approx 1/m_\mathrm{H}$ is not very accurate.

	In this paper we have used both the exact expression for the thermal correction part of the one loop effective action and its usual high temperature expansion. A comparison of the results shows that the use of the exact expression yields a stronger first order transition, as measured by the jump in the order parameter defined by $ \Delta \phi(T_*)=\phi_{\mathrm{true}}(T_*)-\phi_{\mathrm{false}}(T_*)$ and the latent heat released. We have computed all of the important characteristic quantities for the first-order part, including the details of the formation and evolution of bubbles of true vacuum associated with the broken phase, the Higss profile, the duration of the PT, the latent heat, and all of the characteristic temperatures.
	
	We have then explored one of the consequences of the first-order part of EWPT, which is the generation of GWs. In our model, two main sources of GWs are sound waves and MHD turbulence. In particular, we have obtained the energy density spectrum of GWs as a function of frequency and shown that their detection is within the range of Ultimate-DECIGO interferometer for $B^{\mathrm{bg}}_Y\gtrsim 0.45 T_*^2$, for peak frequencies of about $ 0.1-1\,\mathrm{Hz} $. \\\\\\
	
	
	
\end{document}